\documentclass[twocolumn,floatfix,prb,aps,showpacs]{revtex4}
\usepackage{graphicx}
\usepackage{latexsym}
\usepackage{amsmath}
\renewcommand{\vec}[1]{{\bf #1}}
\begin{document}
\bibliographystyle{apsrev}
\title{Fragility of String Orders}
\author{F.~Anfuso and A.~Rosch}
\affiliation{Institute for Theoretical Physics,
University of Cologne, 50937 Cologne, Germany}
%\author{}
%\affiliation{Institute for Theoretical Physics,
%University of Cologne, 50937 Cologne, Germany}
\begin{abstract}
  One-dimensional gapped systems are often characterized by a 'hidden'
  non-local order parameter, the so-called string order.  Due to the
  gap, thermodynamic properties are robust against a weak
  higher-dimensional coupling between such chains or ladders. To the
  contrary, we find that the string order is not stable and decays
  for arbitrary weak inter-chain or inter-ladder coupling.  We
  investigate the vanishing of the order for three different systems:
  spin-one Haldane chains, band insulators, and the transverse field
  Ising model. Using perturbation theory and bosonization, we show
  that the fragility of the string order arises from non-local
  commutation relations between the non-local order parameter and the
  perturbation.
\end{abstract}

\maketitle
% Since the pioneering work of Landau \cite{Landau} that led to the
% theory of 2nd order phase transitions, the concept of order parameter
% has become one of the main paradigms in condensed matter physics. The
% central observation is that when a  symmetry is spontaneously broken,
% certain correlation functions of operators\cite{footnote} $A(x)$ do not decay with distance, 
% \begin{equation}
% \lim_{|\vec{x}-\vec{y}| \to \infty} \langle A(\vec{x}) A(\vec{y})
% \rangle =c  \neq 0
% \label{local}
% \end{equation}
% which allows to define the order parameter $\langle A \rangle=\sqrt{c}$.
% This simple paradigm to describe and discriminate the different phases
% of matter has been successfully applied to a plethora of different
% systems both at finite and zero temperature (among others, classical
% and quantum magnetism, superconductivity, superfluidity, Bose gases).
Since the pioneering work of Landau \cite{Landau} that led to the
theory of 2nd order phase transitions, the concept of order parameter
has become one of the main paradigms in condensed matter physics. 
The central observation is that when a  symmetry is spontaneously
broken and long range order appears into the system,
certain correlation functions of operators \cite{footnote} $A(x)$ do not decay with distance, 
\begin{equation}
\lim_{|\vec{x}-\vec{y}| \to \infty} \langle A(\vec{x}) A(\vec{y})
\rangle =c  \neq 0.
\label{local}
\end{equation} 
and it is possible to define an order parameter as $\langle A
\rangle=\sqrt{c}$. For example, it is known that a classical three-dimensional ferromagnet undergoes a
phase transition, becoming ordered (all the spins point in the
same direction) below a certain
critical
temperature. In this case, the order parameter can be identified with the
spontaneous magnetization $\langle\vec{S}(x)\rangle=m$ 
and the spin-spin correlation function
$\langle\vec{S}(x)\vec{S}(y)\rangle=m^2$ is finite also in the limit $|x-y|\rightarrow\infty$. 
This simple way to describe and discriminate the different phases
of matter has been successfully applied to a plethora of different
systems both at finite and zero temperature. 
Classical and quantum magnetism (with many possible order parameters:
uniform and staggered magnetization, dimerization...), superconductivity (Cooper
pair amplitude), superfluidity (condensate amplitude) are only few of
the many successful applications of the Landau paradigm.

% In the context of low dimensional quantum systems, it has been also
% useful, under some circumstances, to identify `non-local' order
% parameters which cannot be written in the form (\ref{local}). In these
% cases, the order parameter is not directly accessible to any
% experimental probe but can be nevertheless used to mark theoretically
% the boundaries of the different phases. 
% An important example is the so-called ``string-order'' defined by the
% non-local correlation function
% \begin{equation}
% \lim_{|x-y| \to \infty} \langle A(\vec{x}) \left(\prod_{\vec{z} \in S_{\vec{x},\vec{y}}} B(\vec{z})\right) A(\vec{y}) \rangle \neq 0,
% \label{string}
% \end{equation}
% where the operator $\prod_{z \in S_{\vec{x},\vec{y}}} B(\vec{z})$ acts
% on a line (the string) connecting the points $\vec{x}$ and $\vec{y}$.
% One dimensional examples of this family (discussed further below)
% includes the spin-1 chain \cite{Den}  and spin-1/2 ladders
% \cite{Kim}, the transverse-field quantum Ising chain (dual order),
% band- and Mott- Fermionic insulators \cite{Anfuso} and, among Bosonic
% systems, some parameter regime of the Bose-Hubbard
% Hamiltonian~\cite{Altman}.

In the context of low dimensional quantum systems, it has been also
useful, under some circumstances, to identify `non-local' order
parameters which cannot be written in the form (\ref{local}). 
Typically, this is the case of some gapped one-dimensional
Hamiltonians with no local symmetries that are
spontaneously broken and with any two-point correlation function that decays exponentially.
In such systems an hidden long range order can nevertheless be present
and this is encoded in the long distance behavior of certain non-local operators.
Notice that a `non-local' order parameter
is not directly accessible to any
experimental probe but can be equally used to mark theoretically
the boundaries of the different phases. 
In this context, an important example is the so-called ``string-order'' defined by the
non-local correlation function
\begin{equation}
\lim_{|x-y| \to \infty} \langle A(\vec{x}) \left(\prod_{\vec{z} \in S_{\vec{x},\vec{y}}} B(\vec{z})\right) A(\vec{y}) \rangle \neq 0,
\label{string}
\end{equation}
where the operator $\prod_{z \in S_{\vec{x},\vec{y}}} B(\vec{z})$ acts
on a line (the string $S_{\vec{x},\vec{y}}$) connecting the points $\vec{x}$ and $\vec{y}$.
One dimensional examples of this family (discussed further below)
includes the spin-1 chain \cite{Den}  and spin-1/2 ladders
\cite{Kim}, the transverse-field quantum Ising chain (dual order),
band- and Mott- Fermionic insulators \cite{Anfuso} and, among Bosonic
systems, some parameter regime of the Bose-Hubbard
Hamiltonian~\cite{Altman}.

Another important sub-class of these more exotic types of systems is
characterized by the so-called `topological order' (e.g.  fractional
quantum Hall fluids). This concept can be defined by the ground state
degeneracy on non-trivial manifolds \cite{Wen,Senthil,Thooft}. Here we will
only consider systems, as listed above, with unique ground states on a
torus and no topological order.

In this paper, we want to address the following questions: Is the
string order stable against small perturbations? And can it be
generalized from one- to higher-dimensional systems? More precisely,
we will investigate weakly coupled two- or three-dimensional arrays of
one-dimensional systems. In the absence of the higher-dimensional
coupling $\lambda_\perp$ they are characterized by various types of
string order and a finite gap in the spectrum (but no topological
order). Due to this gap, a sufficiently small $\lambda_\perp$ will
never induce any thermodynamic phase transition. Therefore, one might
naively expect that also the order parameter is robust against such a
small perturbation. While this is correct for {\em local} order
parameters, we will show that generically it does not hold for the
non-local string order.

In the following, we will first consider, as a
prototypical example, the disordered phase of the quantum Ising chain in
transverse field characterized by a hidden string order
parameter. Here the divergence of the perturbation theory
 indicates that string order is destroyed by arbitrary
small higher-dimensional coupling $\lambda_\perp$.  We connect this
result -- that also holds for the case of the
spin-1 Haldane chain -- to the band-insulators case where an exact
calculation of the string order-parameter is possible, proving the
absence of string order for any finite $\lambda_\perp$. Using the
language of bosonization, we identify the general mechanism
destabilizing non-local order. In contrast, local order
(e.g. a charge density wave) remains stable.

{\em Transverse field Ising model:}
As a starting point for our discussion, we introduce the Hamiltonian
of the quantum Ising chain 
\begin{equation}
\label{ising}
H=-\sum_i J\sigma^z_i\sigma^z_{i+1}-B\sigma^x_i,
\end{equation}
describing a quantum magnet in a transverse field.
 We will always consider $T=0$ and the ferromagnetic case $J>0$ (
$J<0$ leads nevertheless to completely equivalent physics). The transverse-field Ising chain is a text-book example in the
context of quantum criticality. Upon increasing the tuning
parameter $B/J$, the ground state experiences a quantum
phase transition from a magnetic to a paramagnetic state and the model
can be solved exactly with the
use of the standard Fermionic representation for a quantum spin \cite{Sachdev}.
Interestingly, there is a subtle way to identify the critical value
of $B/J$ that 
exploits a hidden non-local property. In perfect analogy
with the Kramers-Wannier \cite{Kramers} duality transformation for the classical
two-dimensional Ising model, one can introduce the following mapping \cite{Savit}
\begin{eqnarray}
\hat{\mu}_i^x&=&\hat{\sigma}^z_{i+1}\hat{\sigma}^z_{i}\nonumber \\
\hat{\mu}_i^z&=&\prod_{m\leq i}\sigma^x_m,
\end{eqnarray}
that preserves the $SU(2)$ algebra and transforms Eq.~(\ref{ising}) into 
\begin{equation}
\label{isingdual}
H=\sum_i B\mu^z_i\mu^z_{i+1}+J\mu^x_i.
\end{equation}
i.e. one obtains the same  Hamiltonian as in (\ref{ising}) when replacing $J 
\leftrightarrow B, \mu^\alpha_i \leftrightarrow
\sigma^\alpha_i$. Therefore the quantum critical point has to be
located at the self-dual point  $J/B=1$. 

As in the ferromagnetic phase $B<J$, $\langle \sigma^z \rangle$ is
finite, one finds in the disordered phase, $B>J$, that $\langle \mu^z
\rangle$ is finite, or more precisely 
\begin{equation}
\lim_{(j-i)\to \infty} \langle \mu^z_i \mu^z_j \rangle=\lim_{(j-i)\to \infty}
\left\langle \prod_{i<m \le j}\sigma^x_m \right\rangle > 0
\end{equation}
Therefore the disordered phase is characterized by non-local string
order (\ref{string}).
% If we now take an array or just a pair of these quantum Ising chains ,
% we can investigate the fate of the non-local order when a weak
% inter-chain coupling $J_\perp$ is present. For simplicity, we set
% $J=0$ and
% consider the Hamiltonian
% \begin{equation}
% \label{isingcouple}
% H=\sum_i B\sigma^x_{1,i}+B\sigma^x_{2,i}+J_{\perp}\sigma^z_{1,i}\sigma^z_{2,i}
% \end{equation}
% As we have set $J=0$ is trivial to calculate the string order exactly
% \begin{equation}
% \label{ising3d}
% \left\langle \prod_{i<m \le j}\sigma^x_{1,m}
% \right\rangle=\left\langle\sigma^x_{1,1} \right\rangle^{|j-i|} \approx
% e^{-\frac{J_\perp^2}{2 B^2} |j-i|}
% \end{equation}
% where we used that  $\langle \sigma^x
% \rangle\approx 1-J_\perp^2/(2 B^2)$ to leading order in $J_\perp/B$. The
% string order  obviously decays exponentially for any $J_\perp \neq 0$.

If we now take  a pair of these quantum Ising chains,
we can investigate the fate of the non-local order when a weak
inter-chain coupling $J_\perp$ is present. For simplicity, we set
$J=0$ and
consider the Hamiltonian
\begin{equation}
\label{isingcouple}
H=\sum_i B\sigma^x_{1,i}+B\sigma^x_{2,i}+J_{\perp}\sigma^z_{1,i}\sigma^z_{2,i}.
\end{equation}
As we have set $J=0$, the system is a sum of independent two-site
Hamiltonians and is trivial to calculate the string order exactly
\begin{equation}
\label{ising3d}
\left\langle \prod_{i<m \le j}\sigma^x_{1,m}
\right\rangle=\left\langle\sigma^x_{1,1} \right\rangle^{|j-i|} \approx
e^{-\frac{J_\perp^2}{8 B^2} |j-i|}
\end{equation}
where we used that  $\langle \sigma^x
\rangle=2B/\sqrt{4B^2+J^2_{\perp}}\approx 1-J_\perp^2/(8 B^2)$ to leading order in $J_\perp/B$. The
string order, being a product of factors strictly less then 1, decays exponentially for any $J_\perp \neq 0$.

So far, the vanishing of the string order can be an artifact due to
the absence of interactions.
To see that a finite $J<B$ cannot stabilized the order, it is instructive to repeat
the calculation in the dual variables. 
In this language, the order parameter is now local but the coupling $J_\perp$ induces a
non-local term in the Hamiltonian 
\begin{equation}
\label{isingcoupledual}
H=\sum_i B\mu^z_{1,i}\mu^z_{1,i+1}+
B\mu^z_{2,i}\mu^z_{2,i+1}+J_{\perp}(\prod_{m\leq
  i}\mu^x_m)(\prod_{k\leq i}\mu^x_k).
\end{equation}
With standard text-book techniques, we expand the S-matrix to
second order in $J_\perp$ and obtain for the two-point correlation function 
\begin{equation}
\label{perturbcalc}
\langle\mu^z_{1,i} \,\,\, \mu^z_{1,j}\rangle\approx 1-
\frac{J^2_{\perp}|i-j|}{8 B^2}
\end{equation}
consistent with (\ref{ising3d}). Formally, the divergence with $|i-j|$ arises
from the term
\begin{equation}
\int_0^{\infty} dt_1\int_{-\infty}^0dt_2\sum_{k,l}\langle\prod_{m\leq
  k}\mu_m^xe^{-iH_0t_1}\mu^z_i\mu^z_je^{+iH_0t_2}\prod_{s\leq l}\mu^x_s\rangle_0
\label{divergent}
\end{equation}
as a consequence of the non-local commutation relation between the
order parameter and the perturbation
\begin{equation}\label{commut1}
\left[\mu^z_i\,,\,\prod_{m\leq l}\mu^x_m\right]=\left[\mu^z_i\,,\,
  \sigma^z_{l}\right]=i\Bigr(\mu^y_i\prod_{m\leq l, m\ne i}\mu_m^x\Bigl)\Theta (l-i).
\end{equation}
Physically, $\prod_{m\leq l}\mu^x_m$ describes the creation of a
domain wall. As any domain wall created between the points $i$ and $j$
destroys the correlations of $\mu^z_{1,i}$ and $\mu^z_{1,j}$, the
perturbation theory diverges linearly in $|i-j|$.
 From the perturbative argument, it is easy to see
that a finite $J<B$ does not change this picture qualitatively for
$|i-j|$ large compared to the correlation length $\xi \sim
(B-J)^{-1}$. In fact, the interaction term $\sum J\sigma^z_i\sigma^z_{i+1}=\sum J\mu^x_i$ has local commutation
relations with the order parameter and produces only regular
corrections to the perturbation series. These additional non-singular terms  cannot compensate for the
linear divergence induced by the inter-chain coupling.

Even though the long range order vanishes, thermodynamics is
unaffected by $J_\perp$ (up to a small renormalization of the gap):
obviously no quantum phase transition is induced. Only the order
parameter is sensitive to the non-locality of the perturbations. 
In the language of the original variables
$\sigma^\alpha_i$, in contrast, the perturbation was local, the order
parameter non-local, and the same type of divergence arises again due to
their  non-local commutation relations 
(\ref{commut1}).

Finally, notice that in the ferromagnetic phase, $J>B$, the local magnetic order is
stabilized rather than suppressed by $J_\perp$.  While for $B>J$,
$J_\perp$ induces a finite density of virtual `dual' domain wall
fluctuations, the `physical' domain walls are suppressed for $B<J$ by
$J_\perp$. More precisely, they are confined as the
energy of a pair of domain walls with separation $|i-j|$ is
proportional to $|i-j| J_\perp^2/\Delta$ in the ferromagnetic phase \cite{Chaikin}.

{\em Spin 1 chain:} A second major type of non-local order that we want to consider is the so-called string order. 
This was introduced first in 1989 by
Den Nijs and Rommelse \cite{Den} briefly after that Haldane
conjectured the existence of a gap in the spin-1 antiferromagnetic chain
as the hidden order of
the Haldane phase. 
They observed that, even though true N\'eel
order is absent, the ground state has still a form of long range order
characterizing the entanglement of the spins: any site with $S^{z}=\pm
1$ is always
followed by another with $S^{z}=\mp 1$, separated from the first
by a string of $S^{z}=0$ of arbitrary length. This implies that 
the string order parameter 
\begin{equation}
\label{stringorderchain}
SO_{\text{chain}}(i-j)=\Bigl\langle
S^{z}_i\exp\Bigl(i\pi \sum_{l=i+1}^{j-1} S^{z}_l\Bigr) S^{z}_j\Bigr\rangle
\end{equation}
is always finite for $|i-j| \to \infty$ in the Haldane phase. 

 Kennedy and Tasaki\cite{Kennedy1,Kennedy2} observed that this
 non-local order can be understood from the non-local mapping
\begin{eqnarray}
\label{KTmapping}
\tilde{S}^x_j&=&S^x_j\exp\bigl(i\pi\sum_{k=j+1}^{L}S^x_k\bigr)\nonumber\\
\tilde{S}^y_j&=&\exp\bigl(i\pi\sum_{k=1}^{j-1}S^z_k\bigr)S^y_j\exp\bigl(i\pi\sum_{k=j+1}^{L}S^x_k\bigr)\nonumber\\
\tilde{S}^z_j&=&\exp\bigl(i\pi\sum_{k=1}^{j-1}S^z_k\bigr)S^z_j
\end{eqnarray}
as it maps the Heisenberg Hamiltonian with open boundary conditions 
to an effective {\em local}
ferromagnetic Hamiltonian with a manifest $Z_2 \times Z_2$ symmetry
that is fully broken in the ground-state (the origin of the 4-fold ground state
degeneracy for such boundaries is well understood in the AKLT picture \cite{Affleck}
and comes as a consequence of an effective spin 1/2 localized
at each boundary).
The same transformation, applied to the inter-chain
coupling, introduces in the Hamiltonian domain-wall creation
operators similar to the ones of Eq.~(\ref{isingcoupledual}). Therefore
 the absence of the string
order in the presence of a finite $J_\perp$ can be shown as above.
Indeed, the vanishing of the string order for two coupled spin-1
chains was observed numerically by  Todo \emph{et al.} \cite{Todo} using
the quantum Monte Carlo method.

{\em Ladders:} Similar to the spin-1 chain, also gapped spin-$1/2$ ladders
are characterized by a non-local string order \cite{Kim}. In this
case, one can distinguish between two different types
of string orders, $SO_{\rm odd}$ and $SO_{\rm even}$ (see 
Ref.~[\onlinecite{Kim}])
\begin{eqnarray}
SO_{\rm odd}(i-j)&=& -\, \Bigl\langle
(S^{z}_{1,i}+S^{z}_{2,i})\exp\Bigl(i\pi \sum_{l=i+1}^{j-1}
S^{z}_{1,l}\nonumber \\ &&+S^{z}_{2,l}\Bigr)
(S^{z}_{1,j}+S^{z}_{2,j})\Bigr\rangle \label{odd}\\
SO_{\rm even}(i-j)&=&-\, \Bigl\langle
(S^{z}_{1,i+1}+S^{z}_{2,i})\exp\Bigl(i\pi \sum_{l=i+1}^{j-1}
S^{z}_{1,l+1}\nonumber\\ & &+S^{z}_{2,l}\Bigr)
(S^{z}_{1,j+1}+S^{z}_{2,j})\Bigr\rangle.\label{even}
\end{eqnarray}
where the spin-$1$ of Eq.~(\ref{stringorderchain}) is replaced by the
sum of two spin-$1/2$ operators on either the vertical or on the
diagonal rung.

% In a recent paper \cite{Anfuso}, we have shown that the ground state
% of gapped spin-$1/2$ ladders (and of the spin-1 chain) is
% adiabatically connected to the one of an ordinary, non-interacting
% band insulator. Remarkably, the string order (\ref{odd},\ref{even}) turns out to be also
% present in the band insulating phase. For the non-interacting case,
% the string order can be calculated exactly\cite{Anfuso} even for an
% array of such insulators. As shown in Fig.~\ref{2dladder}, the string
% order decays exponentially for arbitrarily weak inter-ladder coupling
% \begin{eqnarray}
% SO \sim e^{-\alpha |i-j|}, \qquad \alpha \propto t_\perp^2
% \end{eqnarray}
% as in Eq.~(\ref{ising3d}). The prefactor $\alpha$ is quadratic in
% the inter-ladder hopping $t_\perp$ (see inset of Fig.~\ref{2dladder}).

In a recent paper \cite{Anfuso}, we have shown that the ground state
of gapped spin-$1/2$ ladders (and of the spin-1 chain) is
adiabatically connected to the one of an ordinary, non-interacting
band insulator. In Ref.~[\onlinecite{Anfuso}], we introduced the
following family of ladder Hamiltonians
\begin{multline}
H=\sum_{i,\alpha,\sigma} t_\alpha a^{\dagger}_{\alpha,i,\sigma}a_{\alpha,
  i+1,\sigma}+h.c.-\frac{U}{2} n_{\alpha,i,\sigma}\\ + \sum_{i,\sigma} t_R a^{\dagger}_{1,i,\sigma}a_{2,i,\sigma}+
t_D a^{\dagger}_{1,i+1,\sigma}
a_{2,i,\sigma}+h.c.\\+ U \sum_{i,\alpha}
n_{\alpha,i,\uparrow}n_{\alpha,i,\downarrow}
+J_R \sum_i{\bf S}_{1,i}{\bf S}_{2,i} 
 \label{hamiltonian} 
\end{multline}
(where $\alpha=1,2$ and $\sigma=\pm$ are the row and spin indices)
whose phase diagram 
includes both Mott ($U,J_R\gg
t_{i}$) and band
insulating phases ($U=J_R=0$). Remarkably, the string order (\ref{odd},\ref{even}) turns out to be finite
also for the band insulator.

\begin{figure}
\includegraphics[width=.50\textwidth,clip]{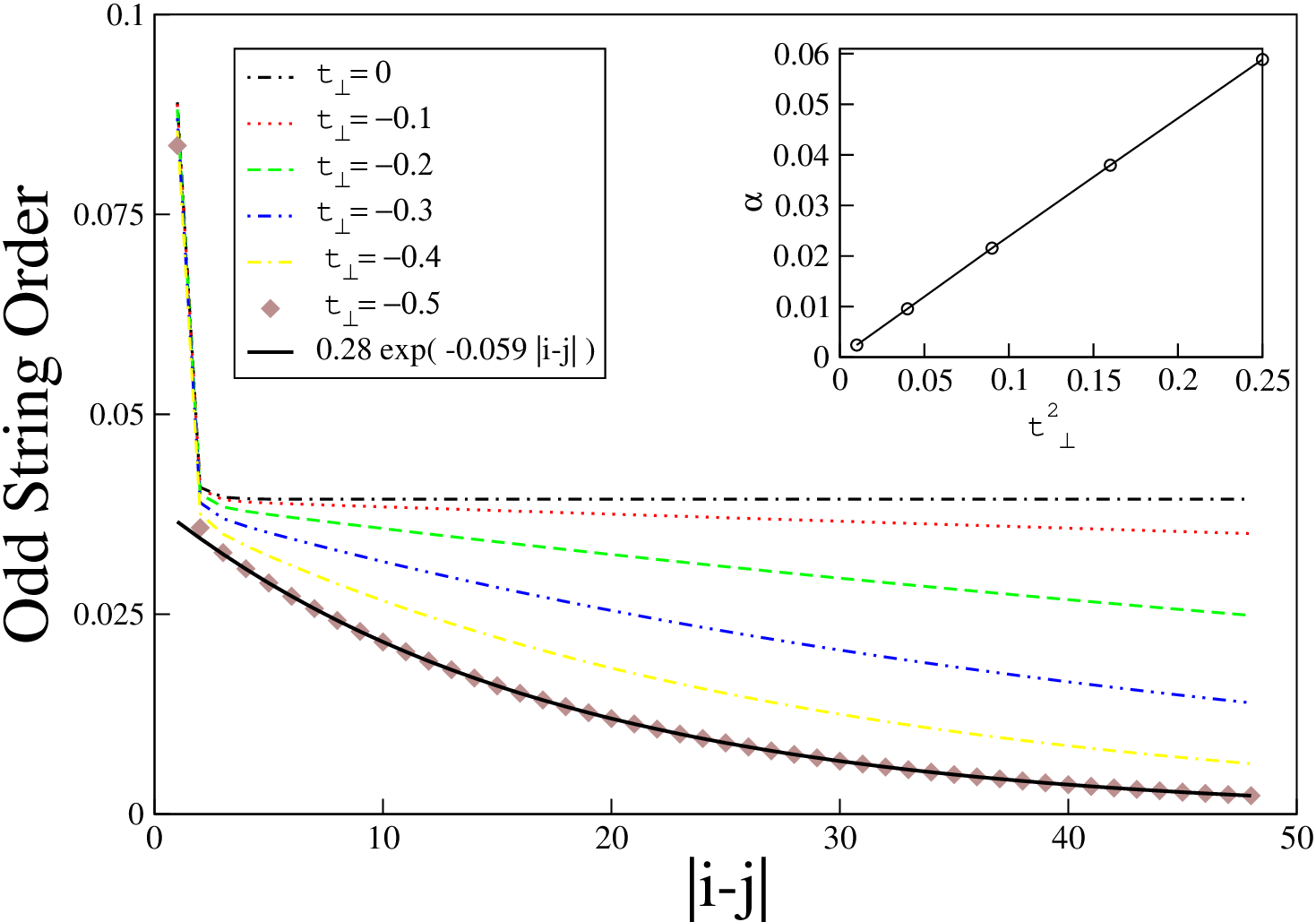}
\caption{(Color online) The odd string order for a two-dimensional
array of weakly coupled non-interacting band insulators decays as
$SO_{\rm odd}\approx e^{-\alpha |i-j|}$. For $t_\perp=-0.5$ we also
show the exponential fit. Inset: $\alpha$ as a function
of $t_{\perp}^2$.
 Both plots are done for the
set of parameters $t_1=0$,
$t_2=-0.6$, $t_R=-1.5$, $t_D=-2$ of the model defined in
Eqns.~(\ref{hamiltonian}, \ref{hamiltonian1}).}
\label{2dladder}
\end{figure}

In the non-interacting case ($U=J_R=0$),
the string order can be calculated exactly\cite{Anfuso} even for an
array of such one-dimensional insulators. As shown in
Fig.~\ref{2dladder},in the presence of an  arbitrarily weak inter-ladder coupling
\begin{equation}
H_\perp=t_\perp \sum_{i,l,\alpha,\sigma} a^{\dagger}_{2,i,l,\sigma}a_{1,i,l+1
  ,\sigma}+h.c. 
\label{hamiltonian1}
\end{equation} 
(where the extra index $l$ labels the ladders),
the string
order decays exponentially 
\begin{eqnarray}
SO \sim e^{-\alpha |i-j|}, \qquad \alpha \propto t_\perp^2
\end{eqnarray}
as in Eq.~(\ref{ising3d}). The prefactor $\alpha$ is quadratic in
the inter-ladder hopping $t_\perp$ (see inset of Fig.~\ref{2dladder}).

%\begin{figure}
%\includegraphics[width=.45\textwidth,clip]{ladderpic.eps}
%\caption{The non-interacting model of Eq.\ref{hamiltonian} with an
%  additional higher dimensional coupling $t_{2s}$.}
%\label{laddermodel}
%\end{figure}

{\em Bosonization:} To identify the precise origin of the fragility of
string order, we now analyze a generic one-dimensional gapped phase
using the language of (Abelian) bosonization. Within this approach,
the relevant degrees of freedom are described by the fields $\Phi$ and
$\Theta$ obeying the non-local commutation relation
\begin{equation}
\bigl[\Phi(x),\Theta(x')\bigr]=i\theta(x-x')
\label{bosecommutation}
\end{equation}
where $\theta(x)$ is the usual $\theta$-function.
Gapped phases
typically arise when one of the Bosonic fields, e.g. $\Phi$, is locked by
a 'relevant' non-linear interaction
\begin{equation}
\label{bosehamilton}
H= \int dx \frac{v_F}{2} \left[ \frac{1}{K} (\partial_x\Theta)^2+ K (\partial_x\Phi)^2\right]
+g\cos(\chi \Phi)
\end{equation}
where $K$ is the Luttinger liquid parameter and $g$ and $\chi$
parameterize the most relevant perturbation of the Luttinger liquid
fixed point. The Hamiltonian has two symmetries: First, it is
invariant under the shift $\Theta \to \Theta+c$ by an arbitrary
constant $c$ reflecting the conservation of the 'charge' $\int
\partial_x \Phi$ (typically either the total
$S^z$ or the total number of electrons). Second, the field $\Phi$ is invariant under a shift
$\Phi \to \Phi+ c_\phi$ by a {\em fixed} number $c_\phi$, reflecting
charge quantization and giving rise to the constraint $\chi=n \, 2
\pi/c_\phi$ with integer $n$. For example, in the case of a
non-interacting band-insulator $c_\phi=2\sqrt{\pi}$, $\chi=\sqrt{4
  \pi}$, $K=1$ and $g$ is
proportional to the $2 k_F$ (where $k_F$ is the Fermi momentum) component of the periodic potential. 

In this language the origin of the string order is easy to
understand. As the field $\Phi$ is locked in one of the minima of the
cosine term, any correlation function of the form
\begin{eqnarray}
O_\gamma(x-y) = \left\langle e^{i \gamma \Phi(x)} e^{-i \gamma \Phi(y)} \right \rangle
\label{gamma}
\end{eqnarray}
does not decay for $|x-y| \to \infty$ and has a finite value
\begin{eqnarray}
\lim_{|x-y| \to \infty} O_\gamma(x-y) = \mu_{SO}^2.
\end{eqnarray}
%has a finite value.

\begin{figure}
\includegraphics[width=.49\textwidth,clip]{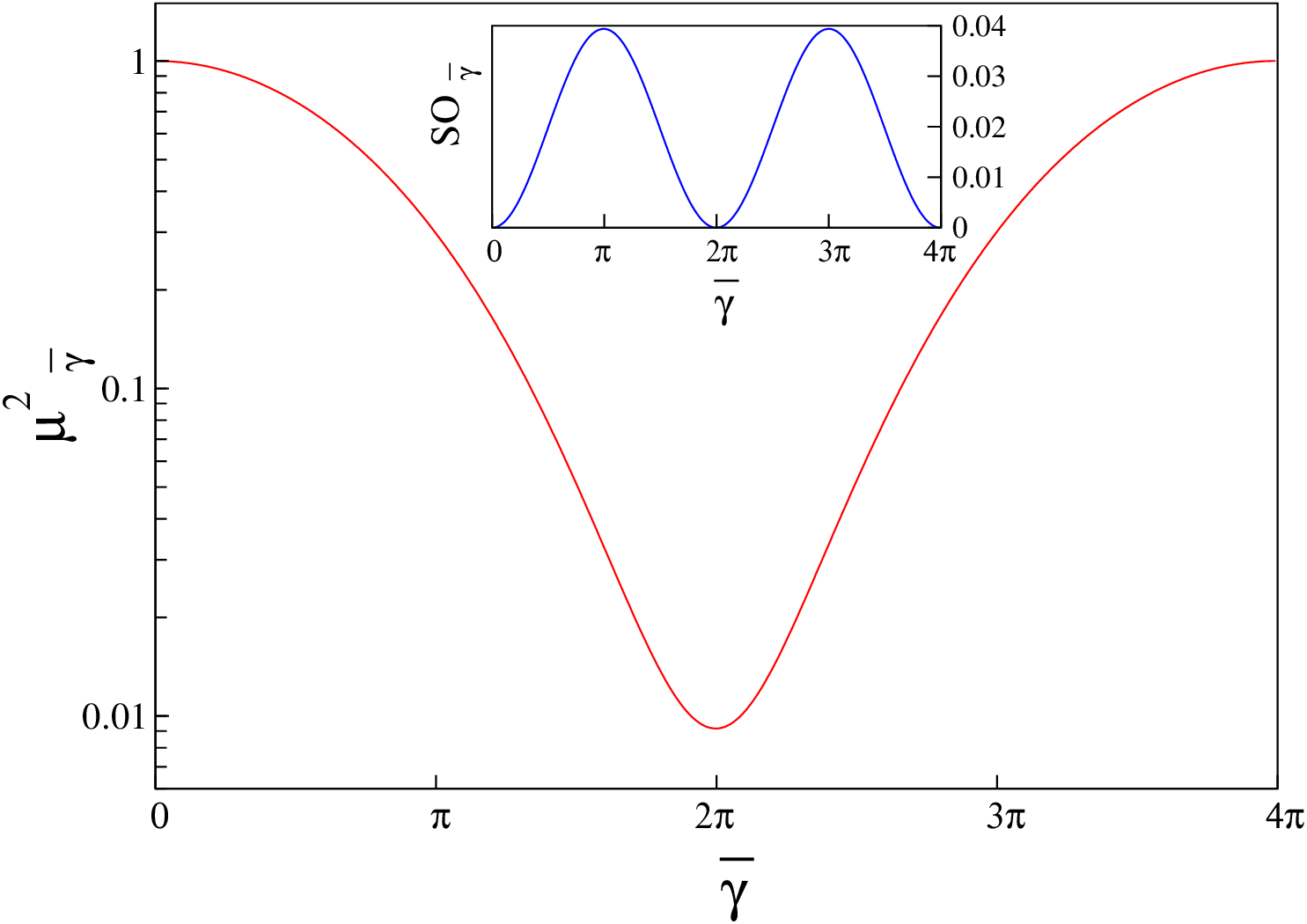}
\caption{(Color online) The string order parameter without including the
  boundary spins defined as $\mu_{\bar\gamma}^2=\lim_{j\to
    \infty} \left \langle 
\exp\left(i \bar\gamma \sum_{l=1}^{j-1}
S^{z}_{1,l}+S^{z}_{2,l}\right)\right\rangle$ for a one-dimensional 
band insulator on a ladder (for the value of the parameters, see
Fig.~\ref{2dladder}) as a function of $\bar\gamma$. By
construction $\mu_{\bar\gamma}=1$ for $\bar\gamma=4 \pi n$. Inset: String order
$SO_{\bar\gamma}=\lim_{j\to \infty} \left \langle 
(S^{z}_{1,j}+S^{z}_{2,j}) \exp\left(i \bar\gamma \sum_{l=1}^{j-1}
S^{z}_{1,l}+S^{z}_{2,l}\right)(S^{z}_{1,0}+S^{z}_{2,0})\right\rangle$  including the
boundary spins. Notice that at $\bar\gamma=2\pi$ the $SO_{2\pi}$ is exactly zero
while the $\mu^2_{2\pi}$ has a small but finite value. This arises
because the operator $e^{i2\pi S_z}$ is independent of the spin configuration.} 
\label{generalizedorder}
\end{figure}

For example, a 'string operator' as in Eq.~(\ref{odd}) is bosonized in
the
continuum limit by\cite{Kim}
\begin{eqnarray}
\exp\Bigl(i \bar\gamma \sum_{l=i+1}^{j-1}
S^{z}_{1,l}+S^{z}_{2,l}\Bigr)
\approx
\exp\Bigl[i\frac{\bar\gamma}{\sqrt{2\pi}}(\Phi_s(x)-\Phi_s(y))\Bigr]\nonumber
\\  \label{spinstring}
\end{eqnarray}
where $\partial_x \Phi_s$ is proportional to the total spin
density. Note that in Eqs.~(\ref{stringorderchain}), (\ref{even}) and 
(\ref{odd}) besides the string operator, also
extra boundary spins at site $i$ and $j$ have been included in the
definition of the string order. However, these terms are not
essential, as discussed in the caption of Fig.~\ref{generalizedorder},
and string order is present as long as $
  O_\gamma(x-y)$ is finite for $|x-y| \to \infty$.

We can make use of this formalism to understand the consequence of
an inter-chain coupling. In the presence of such a perturbation,
the total 'charge' $\int \partial_x \Phi$ on a {\em single} chain is
not conserved any more. This is reflected in  the appearance
of the dual field $\Theta$ in the Hamiltonian (more specifically,
of the exponential $e^{i \beta \Theta(x)}$).
For instance, for chains coupled by single-electron hopping one obtains
\begin{multline}
H_\perp=t_\perp \sum_{l,\sigma=\uparrow/\downarrow}
\Psi^\dagger_{l,\sigma}\Psi_{l+1,\sigma}+h.c.\\
\sim \frac{1}{2\pi a}\sum e^{i\sqrt{\frac{\pi}{2}}(\Phi_{c,l}\pm\Phi_{s,l}+\Theta_{c,l}\pm\Theta_{s,l})}\\\times
e^{-i\sqrt{\frac{\pi}{2}}(\Phi_{c,l+1}\pm\Phi_{s,l+1}+\Theta_{c,l+1}\pm\Theta_{s,l+1})}+... 
\label{hperp}
\end{multline}
where the summation index $l$ spans the different ladders and we introduced the spin and charge Bosonic fields (and their
duals). In the last equality only the
$\Psi^{\dagger}_{L,l,\sigma}\Psi_{L,l+1,\sigma}$ component is shown.
% These hopping terms involve both charge- and spin fields, $\Phi_c,
% \Phi_s$ and their duals $\Theta_c$ and $\Theta_s$. 

% For the following
% discussion all fields ($\Phi_{c,i}, \Phi_{s,i}, \Theta_{c,i},
% \Theta_{s,i \neq l}$) which commute with the string order parameter defined on the chain
% with index $l$,
% $e^{i \gamma (\Phi_{s,l}(x)-\Phi_{s,l}(y))}$, turn out to give only
% finite renormalizations. 
According to the commutation relations  (\ref{bosecommutation}), the
operator $e^{i \beta \Theta(x)}$ increases the $\Phi(x')$ field by $\beta$
for $x'<x$.  In a semi-classical
picture, $e^{i \beta \Theta(x)}$ therefore creates a domain wall at
$x$ by shifting $\Phi$ for $x'<x$
from one minimum of the cosine to another (note that
$\beta$ is always an integer multiple of $2 \pi/\chi$ as a
consequence of charge quantization). 

In analogy to the calculation for the transverse field
Ising model, Eq.~(\ref{divergent}), we can now proceed by calculating the
corrections to the string order parameter  perturbatively in $H_\perp= \int d x
h_\perp(x)$. Here we assume that the string order is calculated on the
chain with index $l=0$ and is defined in terms of the spin field
$\Phi_{s,0}$, $O_\gamma(x-y)= \langle e^{ i \gamma
  (\Phi_{s,0}(x,0)-\Phi_{s,0}(y,0))} \rangle$.
%The operator
%$e^{i \beta \Theta(x)}$ creates a massive excitation with dispersion
%$E_k=\sqrt{\Delta^2+(v k)^2}$, where $\Delta$ is the energy gap and
%$v$ a velocity. Therefore the corresponding correlation function drops
%exponentially on the time scale $1/\Delta$ and the length scale $v/\Delta$.
%\begin{multline}
%\int_{x_1,t_1} \int_{x_2,t_2} \langle T e^{ i \gamma (\Phi(x,0)-\Phi(y,0))}
%e^{i \beta \Theta(x_1,t_1)} e^{-i \beta \Theta(x_2,t_2)}
%\rangle_0.
%\end{multline} 
To second order we obtain
\begin{multline}
\frac{1}{2} \int_{-\infty}^{+\infty}dx_1\, d t_1\int_{-\infty}^{+\infty}dx_2\, d
t_2
\big\langle T e^{ i
  \gamma
  (\Phi_{s,0}(x,0)-\Phi_{s,0}(y,0))}\times\\ h_\perp(x_1,t_1)
h_\perp(x_2,t_2) 
\big\rangle_c
\label{perturbboson}
\end{multline}
Here, $\langle ... \rangle_c$ denotes the connected part of the
correlation function (i.e. $\langle e^{ i \gamma
  (\Phi_{s,0}(x)-\Phi_{s,0}(y))}\rangle \langle h_\perp h_\perp
\rangle$ has been subtracted) and the time-ordering $T$ splits the
integral in the four different contributions $(t_{1}<0<t_{2})$, $(t_{2}<0<t_{1})$
$(t_1,t_2>0)$ and $(t_1,t_2<0)$. As our system is massive,
correlations decay on scale $1/\Delta$, where $\Delta$ is the gap, and
therefore most contributions to (\ref{perturbboson}) are only of order
$(t_\perp/\Delta)^2$. There is, however, one important exception: if a
domain wall is created at time $t_1<0$ and position $x_1$ with $x \ll x_1 \ll y$ and
destroyed at time $t_2>0$, the order parameter changes by $e^{\pm i
  \gamma \beta}$ as at time $t=0$ a domain wall is enclosed between
$x$ and $y$. Here we assume that $|x-y|$, $|x_1-x|$ and $|x_1-y|$ are
much larger than the correlation length $\xi\sim 1/\Delta$. We
therefore obtain (up to corrections of order $\xi/|x-y|$)
\begin{multline}
\frac{1}{2} \int_{-\infty}^{+\infty}dx_1\, d t_1 \,  dx_2\, d
t_2
 (e^{\pm i\gamma \beta \theta(x-x_1)}
e^{\mp i\gamma \beta \theta(y-x_2)}-1)\times \\
 \left \langle  e^{ i
  \gamma
  (\Phi_{s,0}(x)-\Phi_{s,0}(y))} \right\rangle_0  \, \left \langle h_\perp(x_1,t_1)
h_\perp(x_2,t_2) 
\right \rangle\ \\
\approx
 \mu_{SO}^2 |x-y|   (\cos(\gamma \beta)-1)  \times \\ 
 \int_{-\infty}^0 d t_1 \int_{0}^\infty d t_2 \int_{-\infty}^\infty dx' 
\left \langle  h_\perp(x',t_1)
h_\perp(0,t_2) 
\right\rangle \\ \sim
 \mu_{SO}^2 \frac{|x-y|}{a} z (\cos(\gamma \beta)-1) 
\left(\frac{t_\perp}{\Delta}\right)^2 \left(\frac{\xi}{a}\right)^{1-2\eta}
\label{perturbboson2}
\end{multline}
% where $a$ is the lattice spacing and $\eta$ is the scaling
% dimension of $h_\perp$, e.g. $\eta=1$ for weakly coupled band
% insulators and $\eta=1/4$ for $S_z$-$S_z$ coupled Ising chains in a transverse field. The last factor $(\xi/a)^{1-2\eta}$ is only of
% relevance in the asymptotic regime $|x-y|\gg\xi\gg a$, i.e. for a
% gap $\Delta$ small compared to the band width. 

where $\mu^2_{SO}$ is the order parameter in absence of
the perturbation, $a$ is the lattice spacing, $z$ the number of neighboring chains and $\eta$ is the scaling
dimension of $h_\perp$. The last factor $(\xi/a)^{1-2\eta}$ is only of
relevance in the asymptotic regime $|x-y|\gg\xi\gg a$, i.e. for a
gap $\Delta$ small compared to the band width. 

In the cases of the transverse-field Ising model with $J=0$ and
the non-interacting band insulator, we have shown that the
linear correction (\ref{perturbboson2}) resums to an
exponential. Physically, it is clear that this also will happen for
the arbitrary systems considered above: The finite $t_\perp$ induces a
finite density of domain walls and the probability of having no domain
walls between $x$ and $y$ is therefore exponentially small. As the
non-local string correlations are destroyed  by domain walls, the
string order vanishes exponentially,
\begin{equation}
SO \approx \mu_{SO}^2 e^{-\alpha |x-y|}
\end{equation}
where $\alpha$ can be read off from Eq.~(\ref{perturbboson2}) in the
limit of small $t_\perp \ll \Delta$.

\begin{figure}
\includegraphics[width=.50\textwidth,clip]{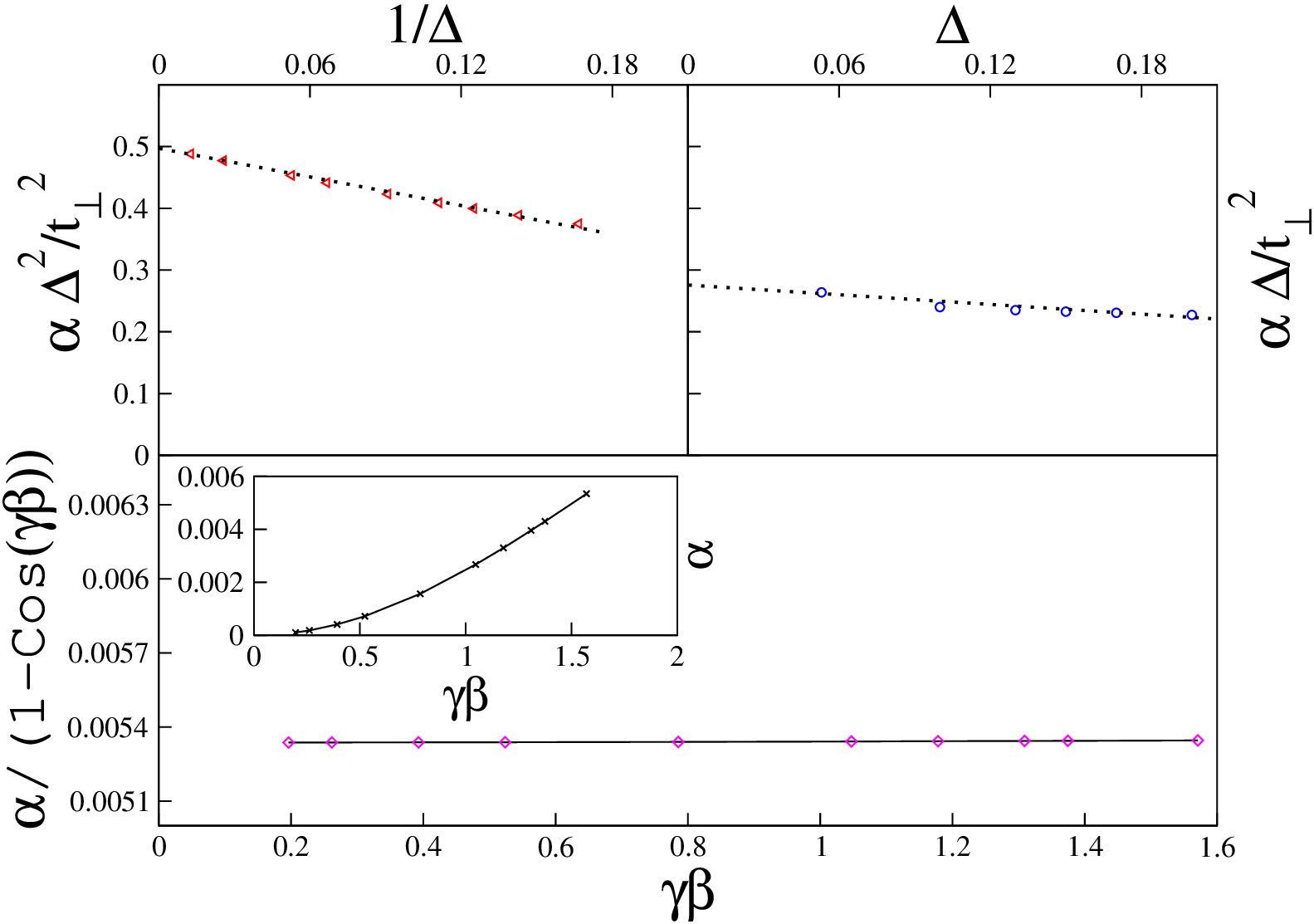}
\caption{(Color online) We test the different limits of Eq.~(\ref{perturbboson2}) in
  the case of an array of band insulators (the Hamiltonian is defined
  in Eqns.~(\ref{hamiltonian},\ref{hamiltonian1})) where we can calculate the
  SO parameters exactly ($SO_{\rm odd/even}\sim e^{-\alpha|i-j|}$).
 Upper left panel: $\alpha$ multiplied by the ratio 
$\Delta^2/t^2_{\perp}$ versus $1/\Delta$ in the limit 
of $\Delta$ much bigger then the band
  width ($\xi\approx a$) for the set of parameters 
$t_1=0$, $t_2=0$, $t_R=-1$, $t_D=[-80,-6]$, $t_{\perp}=-0.2$. The corrections to Eq.~(\ref{perturbboson2})
  are of order $1/\Delta$.
  Upper right panel: 
  $\alpha$ multiplied by the ratio $\Delta/t^2_{\perp}$ versus
  $1/\Delta$ in the scaling limit ($\xi\gg a$) for the set of
  parameters $t_1=0$, $t_2=0$, $t_R=-1$, $t_D=[-1.2,-1.05]$, $t_{\perp}=-0.02$. The corrections to 
Eq.~(\ref{perturbboson2}) are of order $\Delta$. Lower
  panel: $\alpha$ divided by the predicted angular
  dependence $(1-\cos(\gamma\beta))$ for the set of
  parameters $t_1=0$, $t_2=0$, $t_R=-1$, $t_D=-4$, $t_{\perp}=-0.4$ ($\beta=\sqrt{\pi/2}$ is fixed by the low
  energy expression of the inter-chain coupling and we vary
  $\gamma$ in Eq.~(\ref{gamma})). 
  Lower panel inset: the decay exponent versus $\gamma\beta$ for the
  same set of parameters.} 
\label{provaformula}
\end{figure}

In Fig.~\ref{provaformula}, we  compare
Eq.~(\ref{perturbboson2})
with the exact calculation available for the weakly coupled band
insulators ($\eta=1$, $\beta=\sqrt{\pi/2}$). We find 
nice agreement with the expected $\alpha\approx 1/\Delta^2$ and $\alpha\approx 1/\Delta$
behavior both deep in the gapped phase, where $\xi\approx a$ (left upper panel of Fig.~\ref{provaformula})
and close to the critical point where $\xi\gg a$ (right upper panel of
Fig.~\ref{provaformula}). 
Also the angular dependence
is very well described by the factor $(1-\cos(\gamma\beta))$ (see lower panel of Fig.~\ref{provaformula}).

For the case of the quantum Ising chain in transverse field, the $J=0$
case  has already been  considered in Eq.~(\ref{ising3d}). We can
now use
our result to predict the leading behavior of the decay
exponent for $B \gtrsim J$. As the scaling dimension for the
$S_z$$S_z$ coupling between neighboring chains is given \cite{Sachdev} by $\eta=1/4$, it follows from
Eq.~(\ref{perturbboson2}) that the string order 
decays as 
\begin{equation}
SO(x-y)\approx \mu_{SO}^2 e^{-\frac{cJ_{\perp}^2|x-y|}{a \Delta^{5/2}/ \sqrt{J}}}
\end{equation}
where  $c$ is a dimensionless constant of order $1$.

From Eq.~(\ref{perturbboson2}) we can directly infer the conditions of
stability of the string order. in fact, if the relation
\begin{equation}
\gamma \beta=2 \pi n
\label{result}
\end{equation}
holds, the order parameter
commutes with $h_\perp$ and remains finite. For example, in the case
of the band insulator, we have $\beta=\sqrt{\pi/2}$ and the string
order is formally stable for $\bar \gamma= 4 \pi n$ in
Eq.~(\ref{spinstring}). However, in this case the operator
$\exp\left(i \bar\gamma \sum_{l=i+1}^{j-1}
  S^{z}_{1,l}+S^{z}_{2,l}\right)$ becomes trivially the identity and
the correlation function $ \left\langle S^z_{j1} \exp\left(i
    \bar\gamma \sum_{l=i+1}^{j-1}
    S^{z}_{1,l}+S^{z}_{2,l}\right)S^z_{i1}\right \rangle =
\left\langle S^z_{j1} S^z_{i1}\right \rangle$ decays exponentially
even for the purely one-dimensional model (see
Fig.~\ref{generalizedorder}). Therefore no stable order parameter
exists for a band insulator (or the Haldane chain).

%angle:[t1=0,t2=0,tR=-1,tD=-4,tl=-0.4]
%scaling:[t1=0,t2=0,tR=-1,tD=(-1.05,-1.2),tl=-0.02]
%nonscaling[t1=0,t2=0,tR=-1,tD=(-6,-80),tl=-0.2]

As a consistency check, we now analyze a case where the order is
purely local and therefore stable with respect to small
perturbations. For example, spinless Fermions form a charge density
wave for sufficiently strong interactions (of sufficient long
range). Within bosonization such a system is described by \cite{Gogolin}
\begin{equation}
\label{bosehamilton3}
H= \int dx \frac{v_F}{2} \left[ \frac{1}{K} (\partial_x\Theta)^2+ K (\partial_x\Phi)^2\right]
+g\cos(\sqrt{16 \pi} \Phi)
\end{equation}
and for $K>1/2$ the cosine term becomes relevant locking the $\Phi$
field, which implies a spontaneous breaking of the translational
invariance for the underlying lattice model. Equivalently,
(\ref{bosehamilton3}) describes the physics of the XXZ spin-1/2 chain
where the condition $K<\frac{1}{2}$ translates to $J_z>J_{xy}$. In
this case, long-ranged N\'eel order develops in the ground state. In
both cases the relevant order parameter, the staggered component of
the Fermionic density $\Psi_R^\dagger \Psi_L$, is local. Within
bosonization it can be extracted from $\lim_{|x-y|\to \infty} \langle
e^{i 2 \sqrt{\pi} (\Phi(x)-\Phi(y))} \rangle$. A perturbation due to
hopping to the neighboring chain or due to a spin-flip is proportional
to $e^{\pm i \sqrt{\pi} \Theta}$. This implies $\gamma=2 \sqrt{\pi}$,
$\beta=\sqrt{\pi}$ and therefore $\gamma \beta = 2 \pi$, fulfilling
the condition of stability (see Eq.~(\ref{result})): the charge density
wave or the N\`eel order are ``true'' local orders in a gapped system
which are stable with respect to small perturbations.

In conclusion, we have shown that the string order of gapped one-dimensional
systems is very fragile: an arbitrarily small coupling to neighboring
chains or ladders is sufficient for its vanishing. This has to be
contrasted with the behavior of essentially all other ground state
properties which are minimally affected. The finite energy gap
protects them such that a critical coupling is needed to induce a
quantum phase transition.

% One may think about possible generalizations of the string order
% parameter for higher dimensional systems. Indeed, for two coupled
% spin-1 chains, Todo\cite{Todo} suggested a generalization of the string-order
% parameter where $S_{x,y}$ in Eq.~(\ref{string}) includes the spins of
% both chains. While such a construction is possible for a finite number
% of coupled chains or ladders, a direct generalization for a two- or three
% dimensional system (e.g. by including an infinite strip in $S_{x,y}$) 
% is not possible due to the infinite surface of such a structure.

One may think about possible generalizations of the string order
parameter for higher dimensional systems. Indeed, for two coupled
spin-1 chains, Todo\cite{Todo} suggested a generalization of the string-order
parameter (also used in Ref.~[\onlinecite{Kolezhuk}] to characterize the
phases of frustrated spin-1 chains)
\begin{equation}
\tilde{SO}_2=\lim_{|i-j| \to \infty}S^z_{1,i}S^z_{2,i}
e^{\sum_{l=i+1}^{j-1}(S^z_{l,1}+S^z_{l,2})}S^z_{1,j}S^z_{2,j}
\label{Todo}
\end{equation}
\newline
where $S_{x,y}$ of Eq.~(\ref{string}) now includes the spins of both
chains. The stability of $\tilde{SO}_2$ comes from the fact that all
non-local commutation relations of the string-order and the
inter-chain coupling vanish as $\sum S^z_{\alpha i}$ commutes with
$H_\perp$. In a completely analogous way, one can define for any
finite number of chains $N$ a generalized order parameter
$\tilde{SO}_N$ (with or without boundary spins) that is non-zero.
However, a generalization of the above formula for a two- or three-dimensional system 
is not possible. First, $SO_{N}$ vanishes exponentially even in the
{\em absence} of any coupling between the chains. Second, if the
string of Eq.~(\ref{string}) is generalized to a square, a cube or any
other finite subset of spins, this will immediately lead again to
non-local commutation due the presence of 'dangling singlets' at the
boundaries of such structures. 
% Therefore the string order will decay as
% $e^{-\alpha A}$ where $A$ is the area of the boundary.
As for dimensions $d>1$ the surface of any non-local structure of
infinite extension is infinite, we conclude that no direct
extensions of
string order to higher dimensional systems can be deviced
and generalized string orders will decay as
$e^{-\alpha A}$ where $A$ is the area of the boundary \cite{foot}.

An interesting question for the future is the investigation of the
stability of various
types of topological order. For example, Senthil and
Fisher\cite{Senthil} have shown that the ground state degeneracy of
  the deconfined phase of a
  two-dimensional $Z_2$ gauge theory is stable with respect to a small
  inter-layer coupling. Finally, a related problem is  the
  stability of various types of entanglement measures \cite{Verstraete,Venuti}.

We acknowledge useful discussions with J.I.~Cirac, M. Garst,
R.~Moessner,
 E.~M\"uller-Hartmann, A.A.~Nersesyan, A.~Schadschneider, A.M.~Tsvelik and
 J.~Zaanen, J.~Zittartz and, especially, G.I.~Japaridze. We thank for financial support of the DFG under SFB 608.


\begin{thebibliography}{999}
\bibitem{footnote} Here we assume that $\langle A \rangle=0$ in the
  absence of spontaneous symmetry breaking.
\bibitem{foot} Our discussion refers to order parameters defined as a string, area or
volume which begins and ends at two points $i$ and $j$, respectively, such
that the surface of the structure diverges in the limit $|i-j| \to \infty$.
It does not cover, for example, the case of possible generalizations 
involving all the sites in the system.
\end{thebibliography}
\end{document}